\documentclass[a4paper,11pt]{article}
\pdfoutput=1
\usepackage{dcolumn}
\usepackage{bm}
\usepackage[T1]{fontenc}
\usepackage{graphicx}
\usepackage{amssymb,amsmath}
\usepackage{booktabs}
\usepackage{multirow}
\usepackage{cite,color,url}
\usepackage[dvipsnames]{xcolor}
\def\linkcolor{cyan!70!black}

\usepackage[
colorlinks=true
,urlcolor=\linkcolor
,anchorcolor=\linkcolor
,citecolor=\linkcolor
,filecolor=\linkcolor
,linkcolor=\linkcolor
,menucolor=\linkcolor
,linktocpage=true
,pdfproducer=medialab
,pdfa=true
]{hyperref}
\usepackage{tikz}
\usepackage{mathtools}
\usepackage[capitalise]{cleveref}

\usepackage{slashed}
\usepackage{epsfig,psfrag,rotating,soul}
\usepackage{rotfloat}
\usepackage[font={small}]{caption}
\usepackage{colortbl}
\usepackage{soul}
\usepackage{xtab}
\usepackage{listings}

\definecolor{codegreen}{rgb}{0,0.6,0}
\definecolor{codegray}{rgb}{0.5,0.5,0.5}
\definecolor{codepurple}{rgb}{0.58,0,0.82}
\definecolor{backcolour}{rgb}{0.95,0.95,0.92}
\lstdefinestyle{mystyle}{
  backgroundcolor=\color{backcolour},   commentstyle=\color{codegreen},
  keywordstyle=\color{magenta},
  numberstyle=\tiny\color{codegray},
  stringstyle=\color{codepurple},
  basicstyle=\ttfamily\footnotesize,
  breakatwhitespace=false,         
  breaklines=true,                 
  captionpos=b,                    
  keepspaces=true,                 
  numbers=left,                    
  numbersep=5pt,                  
  showspaces=false,                
  showstringspaces=false,
  showtabs=false,                  
  tabsize=2
}
\def\param{{\bf \Theta}}
\def\obs{{\bf X}}

\usepackage[normalem]{ulem}

\newcommand\thefont{\expandafter\string\the\font}

\definecolor{lime}{HTML}{A6CE39}
\DeclareRobustCommand{\orcidicon}{%
    \begin{tikzpicture}
    \draw[lime, fill=lime] (0,0) 
    circle [radius=0.16] 
    node[white] {{\fontfamily{qag}\selectfont \tiny ID}};
    \draw[white, fill=white] (-0.0625,0.095) 
    circle [radius=0.007];
    \end{tikzpicture}
    \hspace{-2mm}
}
\newcommand{\orcid}[1]{\href{https://orcid.org/#1}{\orcidicon}}

\evensidemargin \oddsidemargin

\usepackage{fullpage}
\usepackage{geometry} 
\newcommand{\be}{\begin{equation}}
\newcommand{\ee}{\end{equation}}

\newcommand{\beq}{\begin{equation}} 
\newcommand{\eeq}{\end{equation}} 
\newcommand{\ba}{\begin{array}}  
\newcommand{\ea}{\end{array}} 
\newcommand{\bea}{\begin{eqnarray}}  
\newcommand{\eea}{\end{eqnarray} }  
\newcommand{\bal}{\begin{align}}
\newcommand{\eal}{\end{align}}   
\newcommand{\bi}{\begin{itemize}}  
\newcommand{\ei}{\end{itemize}}  
\newcommand{\ben}{\begin{enumerate}}
\newcommand{\een}{\end{enumerate}}  
\newcommand{\bc}{\begin{center}}
\newcommand{\ec}{\end{center}} 
\newcommand{\bt}{\begin{table}}
\newcommand{\et}{\end{table}}  
\newcommand{\btb}{\begin{tabular}}
\newcommand{\etb}{\end{tabular}}

\allowdisplaybreaks

\let\OLDthebibliography\thebibliography
\renewcommand\thebibliography[1]{
  \OLDthebibliography{#1}
  \setlength{\parskip}{0pt}
  \setlength{\itemsep}{0pt plus 0.3ex}
}

\makeatletter
\newcommand{\github}[1]{%
   \href{#1}{\faGithubSquare}%
}
\makeatother

\begin{document}

\begin{titlepage}

\thispagestyle{empty}

\def\thefootnote{\fnsymbol{footnote}}

\begin{flushright}
IFT-UAM/CSIC-24-116\\
\end{flushright}

\vspace*{1cm}

\begin{center}

\begin{Large}
\textbf{
Bayesian technique to combine independently-trained Machine-Learning models applied to direct dark matter detection
}
\end{Large}

\vspace{1cm}

{\sc
David~Cerde\~no$^{1}$\orcid{0000-0002-7649-1956}%
\footnote{{\tt \href{mailto:davidg.cerdeno@uam.es}{davidg.cerdeno@gmail.com}}}%
, Martin de los Rios$^{1, 2}$\orcid{0000-0003-2190-2196}%
\footnote{{\tt \href{mailto:martin.delosrios@uam.es}{martin.delosrios@uam.es}}}%
, Andres D. Perez$^{1, 2}$\orcid{0000-0002-9391-6047}%
\footnote{\tt \href{mailto:andresd.perez@uam.es}{andresd.perez@uam.es}}%
}

\vspace{0.5truecm}

{\sl
$^1$Instituto de F\' \i sica Te\'orica, IFT-UAM/CSIC, Campus de Cantoblanco, 28049 Madrid, Spain

\vspace*{0.15cm}

$^2$Departamento de F\' \i sica Te\'orica, Universidad Aut\'onoma de Madrid, Campus de Cantoblanco, 28049 Madrid, Spain
}

\vspace*{2mm}

\end{center}

\vspace{0.1cm}

\renewcommand*{\thefootnote}{\arabic{footnote}}
\setcounter{footnote}{0}

\begin{abstract}
\noindent 
We carry out a Bayesian analysis of dark matter (DM) direct detection data to determine particle model parameters using the Truncated Marginal Neural Ratio Estimation (TMNRE) machine learning technique. TMNRE avoids an explicit calculation of the likelihood, which instead is estimated from simulated data, unlike in traditional Markov Chain Monte Carlo (MCMC) algorithms. This considerably speeds up, by several orders of magnitude, the computation of the posterior distributions, which allows to perform the Bayesian analysis of an otherwise computationally prohibitive number of benchmark points. In this article we demonstrate that, in the TMNRE framework, it is possible to include, combine, and remove different datasets in a modular fashion, which is fast and simple as there is no need to re-train the machine learning algorithm or to define a combined likelihood. In order to assess the performance of this method, we consider the case of WIMP DM with spin-dependent and independent interactions with protons and neutrons in a xenon experiment. After validating our results with MCMC, we employ the TMNRE procedure to determine the regions where the DM parameters can be reconstructed. Finally, we present \texttt{CADDENA}, a Python package that implements the modular Bayesian analysis of direct detection experiments described in this work.
\end{abstract}

\end{titlepage}

\section{Introduction}\label{sec:intro}

Dark matter (DM) is arguably one of the most enigmatic substances in our Universe. Present in vast amounts, its gravitational interaction with ordinary matter can account for a wide range of effects observed in astrophysical and cosmological observations~\cite{Bertone:2004pz}. A plausible hypothesis is that DM corresponds to a new fundamental particle, described by an extension of the Standard Model of particle Physics. The properties of such a particle can be very diverse, as are the cosmological mechanisms by which it might have been produced in the early universe. The experimental detection of this new type of matter through other means than gravity might serve to measure the DM parameters (such as its mass and couplings), thereby contributing to its identification.

DM can be searched for directly through the energy deposited after its interaction with the target material of terrestrial experiments. These are often underground detectors, protected from cosmic rays and radioactive backgrounds, and extremely sensitive to the nuclear or electronic recoils induced by DM collisions~\cite{Billard:2021uyg,Cebrian:2022brv}. Originally, the canonical signature corresponded to a nuclear recoil after the elastic scattering of a DM particle~\cite{Goodman:1984dc}. Given the typical mass of the target nuclei employed in this type of detectors and the energy range (keV) that they are sensitive to, this allows to test candidates with electroweak scale interactions and masses of the order of 1~GeV or above. This is the case of a generic weakly interacting massive particle (WIMP), a paradigmatic candidate that could have been produced thermally in the early universe in the right amount to account for the observed DM abundance. More recently, the search has been extended to electronic recoils~\cite{Essig:2011nj,Essig:2022dfa}, which allow to probe much lighter candidates, associated with other production mechanisms.

Despite the tremendous experimental effort of the past decades, there has not been yet any confirmed direct observation of DM particles, leading to stringent upper bounds on the DM scattering cross section~\cite{Billard:2021uyg}. For generic WIMPs, the leading constraints for DM masses above 10~GeV have been obtained by large experiments based on liquid xenon such as LZ~\cite{LZ:2022lsv}, XENONnT~\cite{XENON:2023cxc} and PANDA-X~\cite{PandaX-4T:2021bab} (with liquid argon detectors quickly catching up~\cite{DEAP:2019yzn,Manthos:2023swh}). Smaller experiments, but with a lower energy threshold, using germanium such as SuperCDMS~\cite{SuperCDMS:2022kse,SuperCDMS:2024yiv} and EDELWEISS~\cite{EDELWEISS:2020fxc}, silicon such us SENSEI~\cite{SENSEI:2020dpa} and DAMIC~\cite{DAMIC:2019dcn,DAMIC-M:2023gxo}, or various materials (sapphire, CaWO$_4$, Si, LiAlO$_2$) in CRESST~\cite{CRESST:2024cpr}, have specialised in the search for low-mass WIMPs below 1~GeV, also considering electronic recoil signatures. Moreover, COSINE~\cite{COSINE-100:2021zqh,COSINE-100:2022dvc} and ANAIS~\cite{Coarasa:2024xec} employ sodium iodine to search for an annual modulation in the DM rate, thereby providing an independent test of the DAMA/LIBRA results~\cite{BERNABEI2020103810}. Next generation multi-ton detectors~\cite{DARWIN:2016hyl,DarkSide20k:2020ymr} will allow us to probe new areas of the DM parameter space, thereby raising the hope of a future detection. In addition, direct detection facilities with directional~\cite{Vahsen:2020pzb} or time dependent~\cite{Freese:2012xd,Davis:2014ama} capabilities might play a crucial role in the next decades to gain sensitivity in the vicinity of the so-called neutrino floor or fog~\cite{Billard:2013qya,OHare:2021utq}.

The DM-nuclear and DM-electronic interactions depend on the parameters of the underlying particle model. In the event of a positive detection, these parameters might be reconstructed from the analysis of the deposited energy, where uncertainties in the astrophysical parameters~\cite{Green:2017odb,Benito:2020lgu} that describe the DM halo and in the nuclear form factors~\cite{Klos:2013rwa,Vietze:2014vsa} play an important role. The parameter space of these theories can be large (for example, there are numerous possible DM couplings to ordinary matter in a completely general theory), which makes this a computationally intensive problem~\cite{SuperCDMS:2022crd}. This often leads to degenerate solutions, in which an observation can be explained by several different couplings or a combination of them. A possible way to resolve degenerate solutions, and to better identify the DM mass and couplings, is the combination of data from different experiments~\cite{Pato:2010zk,Cerdeno:2013gqa,Brenner:2022qku}. This is not a trivial problem, as combining data from different experiments usually requires the definition of a combined likelihood.

In this article, we develop a modular procedure for estimating the DM model parameters from DM direct detection data. The algorithm is built upon Simulation Based Inference (SBI) techniques, more specifically, the so called Truncated Marginal Neural Ratio Estimation (TMNRE) \cite{tmnre} method. These methods take advantage of machine learning algorithms~\footnote{Other machine learning techniques applied to DM direct detection can be found in Refs.~\cite{Khosa:2019qgp, Herrero-Garcia:2021goa,Bras:2022ejk,Lopez-Fogliani:2024gzj,Amaral:2024edw} for phenomenological studies, and in Refs.~\cite{Golovatiuk:2020krw,DRIFT:2021uus,Coarasa:2021fpv,LUX:2022vee,Coarasa:2022zak,CRESST:2022qor} for experimental analyses.} to perform a Bayesian estimation of model parameters {\em without the need to assume a given likelihood}. This method has several advantages. First, once the model is trained, it can be applied to new data to obtain the posterior probabilities much faster than traditional methods. Second, the marginal distributions are computed automatically without the need of integrating the full posterior, which considerably speeds up the computation.

We demonstrate that, with this method, it is straightforward to concatenate the results of different trained models (as long as their are trained to infer the same model parameters) in a modular fashion. Thus, it is possible to combine data from different experiments and data representations without the need to re-train algorithms or assume a combined likelihood. In order to illustrate this, we study the case of WIMP DM, with interactions described by a generic non-relativistic effective field theory (NR-EFT)~\cite{Fitzpatrick:2012ix,Anand:2013yka}. We analyse the sensitivity for a near-future xenon detector, based on XENONnT~\cite{XENON:2020kmp}, to measure the DM parameters considering as signal the usual spin-dependent and spin-independent operators. To exemplify the impact on the parameter estimation when combining different datasets we use several data representations: the total rate, the differential rate and the distribution of the signal and background events in the cS1--cS2 plane that represents the intensity of the two signal channels per interaction. Each data representation is trained independently and combined in a simple and fast way with our method. As expected, the inclusion of more complex data improves the results and extends the region where the parameters can be estimated with XENONnT.

This article is organised as follows. In \cref{sec:bayanalysis}, we introduce the TMNRE statistical method, and in \cref{sec:combAnalysis} we demonstrate that the likelihood ratios of different analyses which have been independently trained can be combined without the need to re-train or to define a combine likelihood. In \cref{sec:DM_param}, we summarise the physics of the DM models, and we provide  details on the simulations employed to create the datasets. In \cref{sec:results}, we apply this method to DM parameter reconstruction, analysing simulated data of a near-future direct detection xenon experiment (using XENONnT as a benchmark). Our conclusions are presented in \cref{sec:conclusions}. Additionally, in \cref{sec:mcmc} we describe the traditional sampling method that we use to validate our results, and in \cref{sec:caddena} we present \texttt{CADDENA} a Python implementation of our modular method for estimating the posterior probability distribution, including a tutorial example.

\section{Truncated Marginal Neural Ratio Estimation for Parameter Reconstruction}  
\label{sec:bayanalysis}

The reconstruction of theoretical parameters from a set of experimental results can be done through a Bayesian analysis, where the knowledge on some parameters is updated with the results of new observations. In particular, given new experimental data, $\obs$ (which can be a real number, a vector, or a more complex structure, such as an image), Bayesian analysis estimates the {\em posterior}, $P(\param|\obs)$, of the model parameters, $\param$, by means of the Bayes theorem,
\begin{equation} \label{eq:bayes}
    P(\param | \obs) = \frac{P(\obs|\param) P (\param)}{P(\obs)}\ ,
\end{equation}
where $P(\obs|\param)$ is the likelihood, $P(\param)$ is the prior and $P(\obs)$ is the evidence \footnote{A comprehensive review on Bayesian analysis can be found in Ref.\cite{hastie}.}.

Traditionally, the posterior is determined by exploring the parameter space (which is usually multidimensional) with Markov Chain Monte Carlo (MCMC) \cite{Metropolis,Hastings} or Nested Sampling  \cite{nestedSampling, multinest} techniques, and assuming some functional form for the likelihood (often modelled as a Gaussian or a Poissonian distribution). In these cases, it is mandatory to have a forward model, ${\mathcal F}(\param)$, which, for a given set of parameters, $\param$, computes a synthetic realisation, $\tilde{\obs}$, that can be compared with the real one. In addition, the marginal posterior of one single parameter, $\param_{0}$, can be obtained by marginalising over all the others,
\begin{equation}\label{eq:marginal_integral}
    P(\param_{0}|\obs) = \int \frac{P(\obs|\param)P(\param) \prod_{j} d\param_{j \neq 0}}{P(\obs)}\ .
\end{equation}

Alternatively, when the likelihood function is intractable (either because the functional form is unknown or its computation is too intensive), one can use Simulation Based Inference techniques \cite{sbi}. These avoid an explicit calculation of the likelihood, and replace it by an implicit one, which is estimated from simulated data. Recall in this sense that simulating data $\obs$, given some model parameters $\param$, is equivalent to sampling from the distribution $\obs \sim P(\obs|\param)$ (i.e., the likelihood function). Hence, by studying the distribution of simulated data it is possible to estimate the likelihood function.

Machine learning methods have become a very useful tool to analyse large datasets, aiming to extract as much information as possible from the data itself. In the last years, several techniques have been developed that improve Bayesian inference by analysing the simulations computed with the forward model $\mathcal{F}(\param)$ \cite{tmnre,swyft, wandelt, papamakarios,Amaral:2024edw}. Among the different SBI methods, there are some which are specifically designed to directly estimate the posterior function \cite{wandelt}, the likelihood function \cite{papamakarios} or the likelihood to evidence ratio as presented in Ref.\cite{tmnre}. In this work, we make use of the TMNRE method~\cite{tmnre}, which uses neural networks to estimate the marginal likelihood-to-evidence ratio, $r(\obs,\param)$, defined as
\begin{equation}
    r(\obs,\param) \coloneqq \frac{P(\obs|\param)}{P(\obs)} = \frac{P(\param|\obs)}{P(\param)} = \frac{P(\obs,\param)}{P(\param)P(\obs)}\ .
\end{equation}

Neural networks are computational models inspired by the structure and functioning of the human brain. They consist of interconnected processing units (usually referred to as neurons or nodes) organised into sequential layers. Each neuron of a layer is composed by a non-linear function that takes as input the weighted sum of the outputs of the previous layer neurons. These weights are tuned during the training process to minimise a given loss function\footnote{For a complete review on neural networks, we refer the reader to Refs.\cite{mitchell,murphy}}. It has been proved that such a model can approximate any function that associates an input data with some output data \cite{universal} (if the output data is a real number or vector the network is called {\em regressor} while if the output data is a class the network is called {\em classifier}).

In the TMNRE method, the neural network is a classifier that distinguishes data sampled from the joint distribution $P(\obs,\param)$ (class $k = 1$) from data sampled from the disjoint distribution $P(\param)P(\obs)$ (class $k = 0$). In other words, the output of the network is the probability of a pair $(\obs,\param)$ of being class $k=1$, i.e., $P(k=1|\obs,\param)$. By means of the so-called likelihood ratio-trick \cite{lik_ratio_trick, fast},  $r(\obs,\param)$ can be expressed in terms of the neural network output as follows,
\begin{align}\label{eq:ratiotrick}
    r(\obs,\param) &= \frac{P(\obs,\param)}{P(\param)P(\obs)} = \frac{P(\obs,\param|k=1)}{P(\obs,\param|k=0)} \nonumber \\ \nonumber
    &=\frac{P(\obs,\param, k=1)}{P(\obs,\param, k=0)} \, \frac{P(k=0)}{P(k=1)} \\ \nonumber
    &= \frac{P(k=1|\obs,\param)}{P(k=0|\obs,\param)}  \\ 
    &= \frac{P(k=1|\obs,\param)}{1 - P(k=1|\obs,\param)} \ ,
\end{align}
where in the second line we used the fact that we will train the machine learning models with balanced datasets, hence 
$P(k=0) = P(k=1)$.

An initial dataset (which in our case consists of simulated data) that contains all the relevant information is needed. The realisations of this dataset are pairs of $(\obs,\param)$, sampled from the joint probability distribution $P(\obs,\param) = P(\obs|\param)P(\param)$, i.e., $\obs$ is generated using the parameters $\param$. These realisations are then identified with a dummy variable $k=1$. In addition, a second class ($k=0$) of pairs $(\obs,\param)$ is created by shuffling the values of $\param$ while keeping the same $\obs$ value. This is equivalent to sampling $(\obs,\param)$ from the disjoint distribution $P(\obs)P(\param)$. In other words, the final dataset will consist on pairs $(\obs_{i},\param_{j})$ matched with $k=1$ if $i=j$ and $k=0$ if $i \neq j$. In Table~\ref{tab:swyft_realisations} we show an example on how to create a dataset with $N$ initial realisations. Notice that for the second class ($k=0$) we use the same $N$ original $(\obs,\param)$ pairs but the values of $\param$ are in a random order.

\begin{table}
    \centering  
    \begin{tabular}{|c|c|c|c|}
        \hline
        Sample & $\quad x \quad$ & $\quad \param \quad$ & $\quad k \quad$  \\
        \hline
        \hline
        1 & $x_{1}$ & $\param_{1}$ & \textcolor{blue}{$1$} \\
        2 & $x_{2}$ & $\param_{2}$ & \textcolor{blue}{$1$} \\
        3 & $x_{3}$ & $\param_{3}$ & \textcolor{blue}{$1$} \\
        \vdots &\vdots &\vdots&\vdots\\
        N & $x_{N}$ & $\param_{N}$ & \textcolor{blue}{$1$} \\
         & & & \\
        N+1 & $x_{1}$ & $\param_{8}$ & \textcolor{red}{$0$} \\
        N+2 & $x_{2}$ & $\param_{2}$ & \textcolor{red}{$0$} \\
        N+3 & $x_{3}$ & $\param_{94}$ & \textcolor{red}{$0$} \\
        \vdots &\vdots &\vdots&\vdots\\
        2N & $x_{N}$ & $\param_{17}$ & \textcolor{red}{$0$} \\
        \hline
    \end{tabular}
    \caption{Example of the dataset needed for the TMNRE method. We assign the label $k=1$ to the observations that come from the joint distribution, i.e., if $x_i$ is generated using the parameters $\param_i$. Without generating new data, we create new samples with label $k=0$ by shuffling the values of $\param$, i.e., pairing the observations $x_i$ with the parameters $\param_j$.} 
    \label{tab:swyft_realisations}
\end{table}

Summarising, the neural network takes a pair $(\obs,\param)$ as input, and outputs the probability of such pair to be class $k=1$. As explained above, during the training process, the algorithm tunes the weights of the network to retrieve such probability. Once the algorithm is trained and fixed, one can easily obtain the parameter posterior probability given a new realisation $P(\param|\obs_{\rm obs})$. To compute it, the observation $\obs_{\rm obs}$ has to be paired with values of $\param$. After that, one must feed the neural network with the pairs ${(\obs_{\rm obs},\param_{i})}_{i=1}^{m}$ to obtain the corresponding probability of each pair of being of class $k=1$. As shown before, this output can be easily converted to the likelihood-to-evidence ratio with \cref{eq:ratiotrick}. Finally, multiplying by the prior $P(\param)$ the posterior probability $P(\param|\obs_{\rm obs})$ is obtained.

In practice, all this is done internally in the \texttt{SWYFT} \cite{swyft,fast,swyftweb} package. This software is a Python implementation of the TMNRE method, which directly computes the posterior probability of the parameters of interest.

\subsection{Combined Analysis} \label{sec:combAnalysis}

Bayesian parameter inference is subject to the choice of a prior function. Ideally, the initial prior (before any experiment is performed) must be non-informative in order not to lead the final posterior. For example, depending on the numerical range of the model parameters, it is common to assume uniform priors or log-uniform priors. Once some experimental data is available, the information on the model parameters can be {\em updated} with the posterior probability obtained from the Bayesian analysis of that first dataset. Thus, for any subsequent analysis, the prior is no longer non-informative, but  contains the knowledge obtained in previous experiments. In practice, this is implemented by using the posterior of previous analysis as a prior for the new analysis. For example, the posterior using a first experimental dataset $\obs_{1}$ would be computed as $P(\param|\obs_{1}) = P(\obs_{1}|\param)P(\param) /P(\obs_{1})$, using a non-informative prior $P(\param)$. Given a second experimental dataset $\obs_2$, the new posterior would correspond to $P(\param|\obs_{2}) = P(\obs_{2}|\param)P(\param|\obs_{1})/P(\obs_{2}) $.

This can be more conveniently expressed as the product of the likelihood-to-evidence ratios 
\begin{align}
    P(\param|\obs_{2}) &= P(\obs_{2}|\param)P(\param|\obs_{1})/P(\obs_{2})  \nonumber \\ 
    &= r(\obs_{2},\param)  P(\param|\obs_{1}) \\
    &= r(\obs_{2},\param)  r(\obs_{1},\param) P(\param)\ .  \nonumber
\end{align}
where, $r(\obs_{1},\param)$ and $r(\obs_{2},\param)$ can be obtained from the output of two different TMNRE analyses. Interestingly, realisations of different kinds (for example, total rate, a spectrum, or an image), each of them analysed with their own specific TMNRE algorithm, can also be combined, as long as they have been trained to infer the same parameters.

Furthermore, this method can be generalised to an arbitrary number of steps (each corresponding to the inclusion of a new measurement),
\begin{equation}\label{eq:swyft_posterior}
    P(\param|\obs_{n}) = \prod_{i=1}^{n}r(\obs_{i},\param) P(\param)\ .
\end{equation}

This property makes the method extremely powerful and convenient. Since each dataset can be trained individually to obtain each likelihood-to-evidence ratio $r(\obs_{i},\param)$, the inclusion of new experimental data on a previous Bayesian analysis is simple and fast. Also, removing a given previous dataset $\obs_j$ from a posterior determination is as simple as not including the corresponding $r(\obs_{j},\param)$ in \cref{eq:swyft_posterior}.

\section{Application to direct detection experiments} 
\label{sec:DM_param}

In order to study the potential of TMNRE method and the strategy outlined in \cref{sec:combAnalysis} applied to direct detection data, we have analysed several examples with synthetic data corresponding to a future observation of DM and then attempted to reconstruct the parameters of the DM model (its mass and couplings). We have made some simplifications: we only considered one NR-EFT operator at a time and we did not include uncertainties in either the astrophysical parameters~\cite{Green:2017odb,Benito:2020lgu} that describe the DM halo or in the nuclear form factors~\cite{Klos:2013rwa,Vietze:2014vsa}. We used the Standard Halo Model as an established benchmark and form factors as incorporated in the public code \texttt{WimPyDD} ~\cite{Jeong:2021bpl}.

For concreteness, we focused our analysis on xenon detectors, although the method is applicable to all other experiments. We have taken XENONnT as a paradigmatic example, and we used the public tools provided by the collaboration to determine the performance for a total exposure of $20$~ton~yr. We have employed three different data representations: the total number of detected events, the number of events per recoil energy bin, and the full cS1--cS2 plane. This has a two-fold goal: first, it allows us to explore the difference between considering only total number of detected events (a simple analysis that can be found in the literature) vs using more complete information, and second, it provides an explicit example of a combined analysis that can be easily done within the TMNRE Bayesian scheme. Then, we trained different \texttt{SWYFT} algorithms for each data representation and combined their individual results for the joint analysis.

All the material necessary to reproduce the results of this manuscript can be found on github~\footnote{\href{https://github.com/Martindelosrios/CADDENA}{https://github.com/Martindelosrios/CADDENA}}, where our new publicly available Python library for performing a modular Bayesian analysis of direct detection experiments can be downloaded. This package, called \texttt{CADDENA}, can be easily installed in any Python environment, it is simple to implement, and is not computationally intensive since pre-trained neural network models and simulation data are included and ready to use. For more details about \texttt{CADDENA} main functions, see \cref{sec:caddena}.

\subsection{DM non-relativistic Effective Field Theory} \label{sec:eft}

The DM-nucleon scattering can be described within the framework of non-relativistic effective field theory, since DM particles are expected to have non-relativistic velocities in the Solar System. The number of viable operators depends on the spin of the DM particle, but is in general very large; e.g., for fermionic DM one has fourteen possible interactions with either protons or neutrons. The effective interaction Lagrangian (for spin $1/2$ DM particles) can be expressed as~\cite{Fitzpatrick:2012ix,Anand:2013yka}:
\begin{equation}
    \mathcal{L}_{\rm EFT} = \sum_{\tau} \sum_{i} c_i^{\tau} \mathcal{O}_i \bar{\chi} \chi \bar{\tau} \tau,
\end{equation}
where the index $i$ represents the sum over all possible operators, $\mathcal{O}_i$, that describe the DM-nucleon interactions, and $c_i^{\tau}$ are the coupling coefficients. The index $\tau$ indicates the interaction basis, either nucleus or isospin, related by,
\begin{equation}
\begin{aligned}
    c_i^0 &= \frac{1}{2}(c_i^p + c_i^n) = A_i \sin{\theta_i},\\
    c_i^1 &= \frac{1}{2}(c_i^p - c_i^n) = A_i \cos{\theta_i},
\end{aligned}
\label{eq:NREFTcouplings}
\end{equation}
with $c_i^0$ ($c_i^1$) the isoscalar (isovector) interaction, and $c_i^p$ ($c_i^n$) the proton (neutron) interaction coupling. On the right-hand side of \cref{eq:NREFTcouplings} we have introduced a change of variables to describe the couplings in polar coordinates with an amplitude, $A_i$, and a phase, $\theta_i$. The shape of the differential rate is independent on $A_i$, being only determined by $\theta_i$ and the DM mass, $m_{\chi}$.

The full set of non-relativistic operators can be found in Ref.~\cite{Anand:2013yka}. For illustrative purposes,  in this work we have limited the analysis to the $\mathcal{O}_1$ and $\mathcal{O}_4$ operators, corresponding to the usual (momentum and velocity independent) spin-independent (SI) and spin-dependent (SD) interactions, respectively, however, our method can be applied to any generic set of NR-EFT operators. In this case, the amplitudes $A_1$ and $A_4$ are related to the spin-independent and spin-dependent contributions to the DM-nucleon scattering cross-section by,
\begin{equation}
\begin{aligned}
    \sigma^{\textit{SI}} &= \frac{A_1^2 \; \mu^2_{\chi\mathcal{N}}}{\pi},\\
    \sigma^{\textit{SD}} &= \frac{3}{16}\frac{A_4^2 \; \mu^2_{\chi\mathcal{N}}}{\pi},
\end{aligned}
\end{equation}
where $\mu_{\chi\mathcal{N}}$ is the reduced mass of dark matter-nucleon system, with $\mathcal{N}=p,n$. Replacing $A_1$ ($A_4$) by $c_1^{\mathcal{N}}$ ($c_4^{\mathcal{N}}$), we obtain the usual expressions that relate proton and neutron couplings with SI (SD) cross sections when only isoscalar interactions are allowed.

\subsection{Preparation of the datasets} \label{sec:simulations}

We have considered the characteristics of the XENONnT detector as benchmark of a liquid xenon experiment, given its exceptional capabilities for rare event detection and extensive current and near-future research program. XENONnT is a liquid xenon dual-phase time projection chamber (TPC) experiment located at an underground research facility at the Laboratori Nazionali del Gran Sasso. The liquid xenon TPC consists of a central liquid xenon volume surrounded by light reflectors, with two arrays of photomultiplier tubes (PMTs) arranged on the top and bottom part of the TPC to detect light signals.

The interaction of an incident particle on the liquid xenon can induce two types of events: nuclear recoils (NRs) and electronic recoils (ERs). The deposited energy can lead to both the emission of a prompt scintillation light through de-excitation of Xe2 dimers and the release of electrons via atomic ionisation. Scintillation light is detected almost immediately by the top and bottom PMTs, and it is referred to as S1 signal. The ionisation electrons drift upwards, by means of an electric field applied across the active target, and are extracted by a stronger  field into a gaseous xenon phase, located between the liquid xenon phase and the uppermost PMT. These produce a proportional scintillation light that is detected by the photosensors and is denoted S2 signal. The position of the interaction vertex is inferred from the S2 signal pattern on the top PMT array ($x-y$ position), and from the time delay between the S1 and S2 signals, due to the drift time of the ionisation electrons ($z$ position). Both the S1 and S2 signals must be corrected due to position dependent effects modelled in the simulator, such as light collection efficiency, inhomogeneous electroluminescence amplification, electron lifetime and extraction efficiency during the drift towards the upper part of the detector. The corrected signals are renamed cS1 and cS2.

We have taken into account five different background components: ER interactions, surface events, radiogenic neutrons, coherent elastic neutrino-nucleus scattering (CE$\nu$NS) and accidental coincidences (ACs). The background is dominated by ER interactions. The main contribution comes from $\beta$-decay processes in the decay chain of radioactive radon, $^{222}$Rn, which is emanated from detector materials resulting in its presence in the entire active volume. Another importance sources of ER interactions come from the decay chain of $^{85}$Kr naturally present in the xenon target, and from unstable xenon isotopes, all distributed uniformly in the active detector. ER interactions also include the elastic scattering of solar neutrinos (E$\nu$ES) off atomic electrons of the liquid xenon target. Surface or wall events are also ER interactions originating from the TPC walls and can be reduced by fiducialisation. The NR background is dominated by radiogenic neutrons produced through spontaneous fission or ($\alpha$, $n$) reactions in detector materials. Solar neutrino interactions, predominantly $^8B$ and $hep$ neutrinos, also contribute to the NR background through CE$\nu$NS. Finally, the background denoted accidental coincidences (ACs) is the contribution due to randomly paired S1-S2 signals.

In order to simulate different interaction types in a direct detection experiment, we modified a software originally created to simulate
XENON1T data, developed by the XENON Collaboration~\cite{PaX,laidbax,blueice}, to match XENONnT specifications~\cite{XENON:2020kmp,XENON:2023cxc} (size of the TPC, fiducial mass, drift field, electron lifetime, photon detection probability $g_1$, effective charge gain $g_2$, background levels, and a projected exposure of 20~ton~yr). To model the backgrounds mentioned above, we used the templates included in the package, except the solar neutrino contribution that we replaced with \texttt{SNuDD}~\cite{Amaral:2023tbs}, a Python package with accurate computations of solar neutrino spectra and scattering rates at direct detection (DD) experiments in the presence, or not, of non-standard neutrino interactions. Notice that these background templates are not the ones that the XENON collaboration employs for their analysis, however they are very useful to probe the robustness and reach of the proposed method in the current manuscript. This software is one of the few public tools that allow to obtain direct detection raw data. We have validated the introduced modifications by comparing the expected number of background events per channel, the expected number of signal events for some DM models, and the exclusion curves with XENON publications~\cite{XENON:2020kmp,XENON:2023cxc} by performing a frequentist analysis. It is important to highlight that the method proposed in this work is independent of the simulator or particular direct detection experiment. A more detailed description of the backgrounds or replacing the detector response/simulator can be easily implemented, however, we do not expect significant changes in our main conclusions.

In order to generate the DM signal, we must provide the theoretical DM differential rate to the XENON simulator, i.e. the expected number of events per recoil energy. To do this, we used \texttt{WimPyDD}~\cite{Jeong:2021bpl}, a code that calculates accurate predictions of nuclear scattering expected rates and spectra for direct detection experiments, for any given target material, NR-EFT operator and parameters, $m_{\chi}$, $\sigma$ and $\theta$. The transferred energy is selected following the expected energy distribution probability, while the propagation and final simulated signal also take into account propagation and efficiency effects (in electron propagation in the xenon medium, extraction, PMT response, light collection), resulting in the stochasticity of the final data.

For each set of initial parameters, we computed the expected DM signal in a 20~ton~yr XENONnT-like experiment, and we simulated as many observations, or pseudo experiments, as needed. In this work we compared three data representations: the total number of detected events, the number of events per recoil energy bin, and the distribution of the events in the cS1--cS2 plane. Although the training of the algorithms is more costly for the latter, this drawback is not present during its application.

In \cref{tab:numberofevents}, we present the expected number of events for the different interaction sources, for a benchmark point with $m_{\chi}=50$ GeV, $\sigma^{SI}=5\times10^{-47}$ cm$^2$, and $\theta=\pi/2$. Notice that the expected number of signal and background events are compatible with the projections for an exposure of 20~ton~yr in Refs.~\cite{XENON:2020kmp,XENON:2023cxc}. Although we obtain a higher contribution for CE$\nu$NS, this can be due to a more accurate computation of the neutrino flux performed with \texttt{SNuDD}. For comparison, we also show the results of a particular pseudo experiment, which takes statistical fluctuations into account.

\begin{table}[t!]
    \centering
    \begin{tabular}{c|c|c}
  \hline
         & Expected number of events & Pseudo experiment \\
    \hline
        ER & 2482 & 2431 \\
        Surface events & 257.2 & 249 \\
        Radiogenic neutrons & 16.12 & 20 \\
        CE$\nu$NS & 12.16 & 12 \\
        Accidental coincidences & 79.00 & 78\\
        DM & 190.9 & 196 \\
        \hline
    \end{tabular}
    \caption{Expected number of events for each model component and number of events for a particular realisation, or pseudo experiment. The DM model corresponds to an $\mathcal{O}_1$ EFT operator with $m_{\chi}=50$ GeV, $\sigma^{SI}=5\times10^{-47}$ cm$^2$, $\theta=\pi/2$, corresponding to the usual SI interaction.}
    \label{tab:numberofevents}
\end{table}

\begin{figure}[t!]
    \centering
    \includegraphics[width=0.8\textwidth]{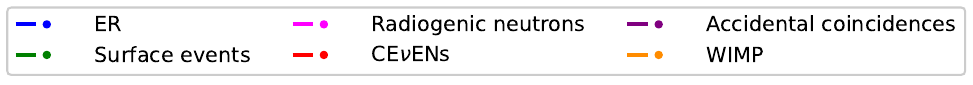}
    \includegraphics[width=0.48\textwidth]{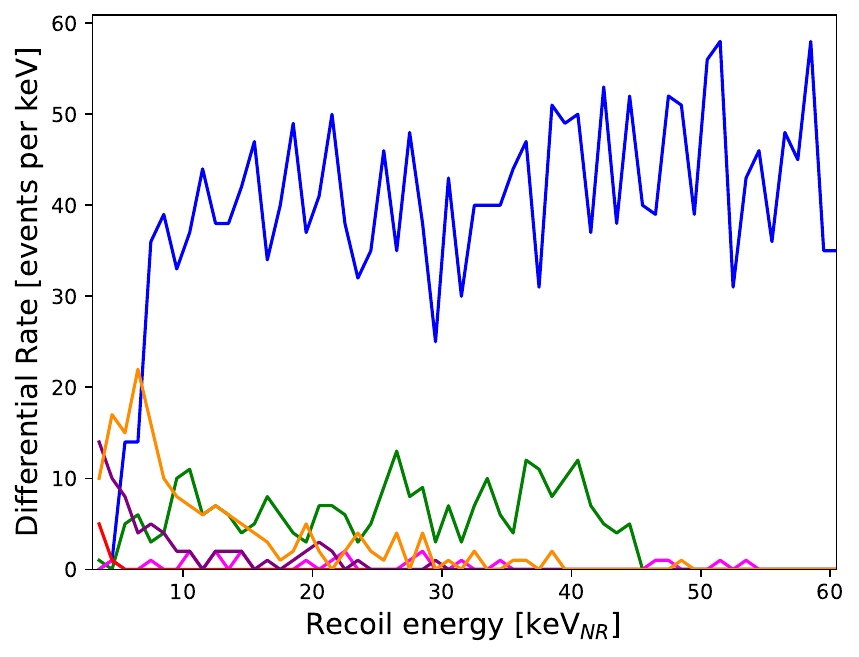}
    \includegraphics[width=0.5\textwidth]{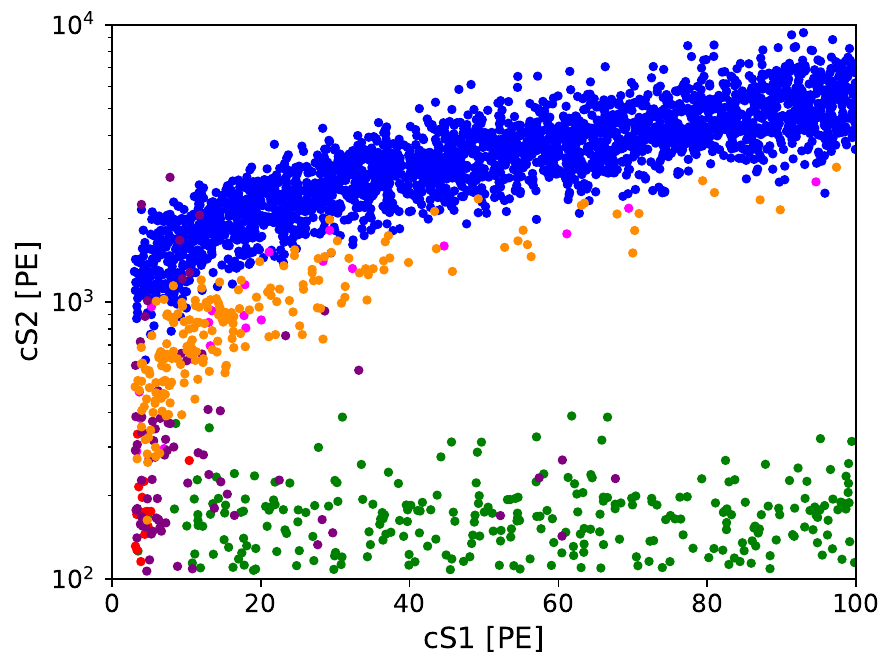}
    \caption{Differential rate (left panel) and distribution of the events (right panel) in the cS1--cS2 space for the same pseudo experiment presented in Table~\ref{tab:numberofevents}.}
    \label{fig:exampledata}
\end{figure}

For the same pseudo experiment of \cref{tab:numberofevents}, the left panel of \cref{fig:exampledata} shows the differential rate for each model component, in an energy range of $(3,\, 61)$~keV$_{\rm NR}$ and a bin width of 1~keV$_{\rm NR}$. For reference, the region of interest in the first DM search with nuclear recoils in XENONnT was $(3.3, 60.5)$~keV$_{\rm NR}$ \cite{XENON:2023cxc}. To construct the differential rate, we count the number of events between isoenergy curves of NR energy in the cS1--cS2 plane. The right panel of \cref{fig:exampledata} shows the distributions of the pseudo-experiment events in the cS1--cS2 space. For cS1, we considered a range of $(3,\, 100)$~PE with 96 bins, and for cS2 a range of $(100, 10000)$~PE with 70 bins. We should clarify that although these panels depict each model component in different colours, this information is not used to train the machine learning classifiers and to generate the posteriors (since these are experimentally indistinguishable).

Notice that we do not apply any cut to separate NR and ER interactions and therefore our study differs from the usual NR data analysis in these detectors. For this reason, the data is largely dominated by the ER background (as evident for example on the left panel of \cref{fig:exampledata}) and the results are not directly comparable to those obtained by the collaborations. Nevertheless, we expect that the TMNRE method can {\em learn} to distinguish the DM NR signal from the rest of the backgrounds in this cS1--cS2 parameter space.
Among the considered data representations, we anticipate that our analysis using cS1--cS2 would yield significantly more accurate results (the advantage of using the whole information of the cS1--cS2 plane was already put forward in Ref.~\cite{Davis:2012hn}, but applied to a standard likelihood analysis).  Thus, analysing this plane without imposing any cuts makes this methodology more general and model independent.

\subsection{Training of the machine learning algorithms}
\label{sec:training}

For the training phase we have generated two datasets (one assuming only $\mathcal{O}_{1}$ interactions and other assuming only $\mathcal{O}_{4}$ type interactions) of 20k realisations (including the corresponding background) sampling random values of $[m_{\chi}, \sigma, \theta]$. We considered a linear prior for $m_{\chi} \in [6,1000]$ GeV, $\theta \in [-\pi/2, \pi/2]$, and a log prior for $\sigma^{SI} \in [10^{-50}, 10^{-43}]$ cm$^2$ for $\mathcal{O}_{1}$, and $\sigma^{SD} \in [10^{-43}, 10^{-35}]$ cm$^2$ for $\mathcal{O}_{4}$.

In order to avoid overfitting, each dataset was split into three mutually independent ones, namely: a training set, a validation set (used to monitor the performance of the neural network after each training epoch), and a test set (to estimate the final performance of the algorithm). For each data representation (total rate, differential rate and cS1--cS2 plane), we have trained two machine learning networks: one for the data including only $\mathcal{O}_{1}$ interactions, and one for the data with only $\mathcal{O}_{4}$ interactions. Hence, in total we have six different TMNRE algorithms.

As is common in machine learning techniques, in order to achieve a better performance all the data was normalised between 0 and 1. In addition, as explained in Ref.~\cite{fast}, when dealing with complex (or high-dimensional) data it is beneficial to perform a data compression before the actual ratio estimation. 
In our case, for the differential rate and cS1--cS2 representations, we added an extra neural network to reduce the dimensionality of the problem.
For the differential rate analysis, we add a dense neural network composed of 5 hidden layers of 500, 1000, 500, 50 and 5 neurons with a ReLu activation function.
For the cS1--cS2 case, we used a CNN composed of a 2D convolutional block with 10 $5 \times 5$ filters followed by a max pooling with 2 and ReLu activation function, followed by a second convolutional block composed of 20 $5 \times 5$ filters, again followed by a max pooling layer. Finally, we add a flatten layer followed by 3 dense hidden layers of 80, 50 and 10 units respectively. These networks reduce the input parameters to 5 and 10, and we have checked that increasing the final number of parameters does not improve the performance.
Also, we have analysed how the validation loss after training varies with the number of training realizations. We found that the performance already stabilize when training with around $\sim 15$k realizations.
Furthermore, we have found that the final results are robust to the choice of architectures of the compression networks.

We should emphasise that the machine learning algorithms trained with different data representations but within the same operator $\mathcal{O}_{1}$ ($\mathcal{O}_{4}$) data are designed to analyse the same parameters space, and hence, they can be combined. Notice, however, that one can not combine algorithms trained on different parameter spaces.

\section{Validation and reconstruction of DM parameters}\label{sec:results}

In order to test the performance of the TMNRE method, with the combination technique introduced in \cref{sec:combAnalysis}, we have first applied it to the reconstruction of DM parameters from synthetic data for some selected benchmark points, comparing the results of the posterior distributions with those obtained by the traditional MCMC approach.

\begin{figure}[t!]
    \centering
    \includegraphics[width=\textwidth]{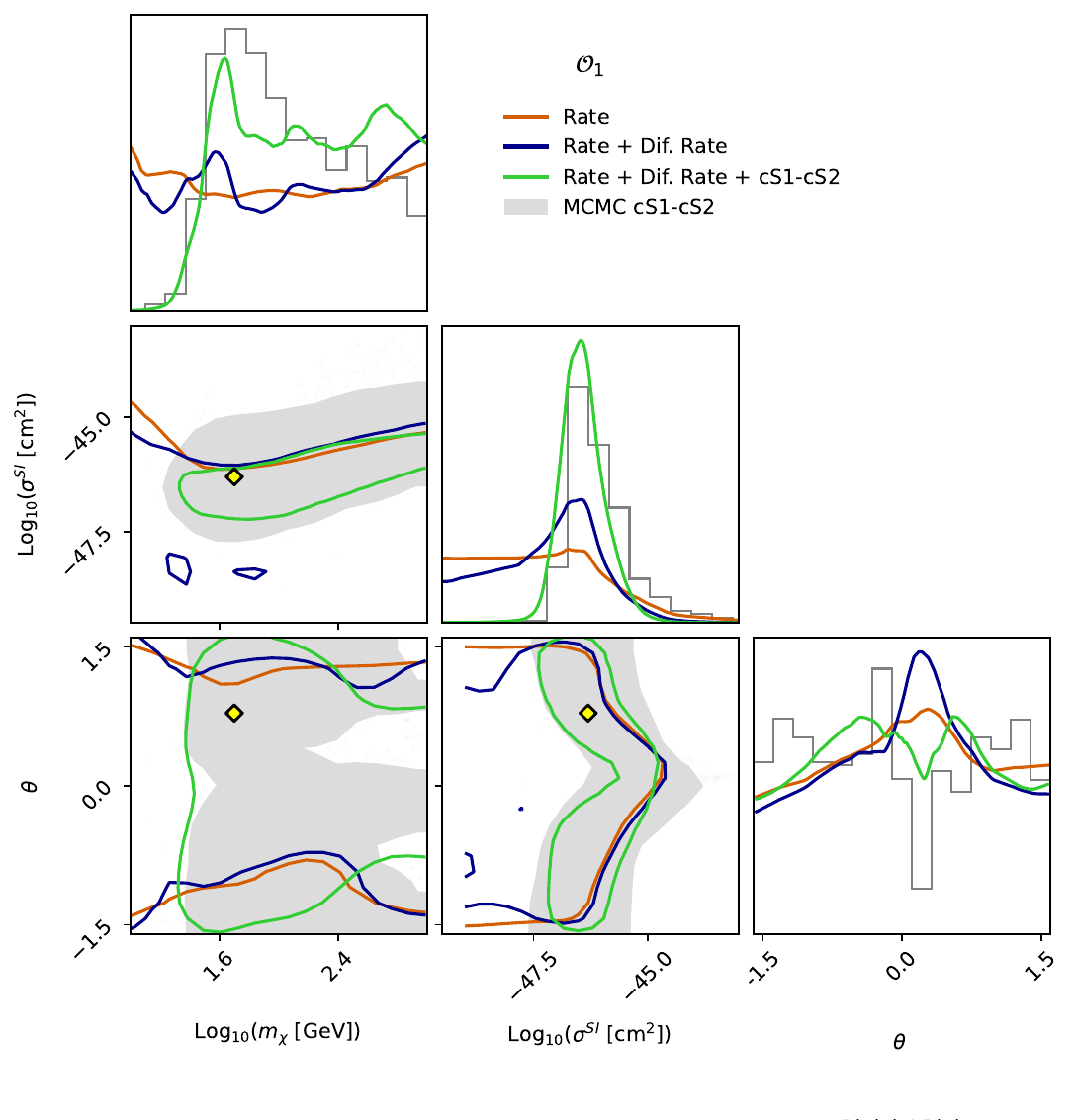}
    \caption{$90\%$ probability regions of the posterior distributions for a dark matter particle with an $\mathcal{O}_{1}$ interaction, $m_{\chi} = 50$ GeV, $\sigma^{SI} = 2\times10^{-47}$ cm$^2$, and $\theta = \pi/4$. 
    In the diagonal we present the $1$D mass, cross-section, and phase marginal posterior distributions, while in the lower left panels we show the $2$D marginal posterior distributions. 
    In orange are presented the results obtained analysing the total number rate, in blue the results when adding also the differential rate information and in green the results when adding also the full cS1--cS2 expected signal.
    The grey areas correspond to the posteriors from an MCMC analysis.
    }
    \label{fig:2d_posterior}
\end{figure}

As a first example, we have analysed the data expected for a benchmark point with $\mathcal{O}_{1}$ operator,  $m_{\chi} = 50$ GeV, $\sigma^{SI} = 2\times10^{-47}$ cm$^2$, and $\theta = \pi/4$. In \cref{fig:2d_posterior} we show the $90\%$ probability regions of the posterior distributions of the analysed parameters when using the expected total rate (orange lines), the total rate combined with the differential rate (blue lines) and the total rate combined with the differential rate plus the full cS1--cS2 image (green lines). For comparison, and to validate the analysis performed with TMNRE, we show in grey the $90\%$ contours obtained with MCMC (implemented in \texttt{MultiNest}~\cite{Feroz:2007kg,Feroz:2008xx,Feroz:2013hea}) using the full cS1--cS2 information (see Appendix~\ref{sec:mcmc} for further details). For completeness, we include the $1$D marginal posterior distributions in the diagonal panels.

The analyses are consistent with the true values of the benchmark point (shown as a black and yellow diamond). As expected, the addition of new data is crucial to put better constraints on the different parameters. For example, analysing only the total rate it is not enough for determining the mass of such hypothetical DM particle (uppermost panel) because for every mass value considered a very similar total number of events can be achieved. This picture changes drastically when including the differential rate and the full cS1--cS2 information, allowing not only to establish a lower bound for the mass but to delimit the cross section. However, the degeneracy in $\theta$ generally can not be resolved for any benchmark point with a single target. Including information from other experiments with complementary target nuclei, i.e., with different suppression phases (for xenon $\theta\simeq0.17$), would help constraint this parameter.

It is also important to remark that, although the results obtained with TMNRE and with traditional MCMC are consistent\footnote{We have checked that the results are also consistent using the total rate and the differential rate (see Appendix~\ref{sec:mcmc}).}, there are some differences.
These discrepancies are mainly due to the fact that, as explained in \cref{sec:bayanalysis}, for the MCMC analysis we need to assume a likelihood, while in TMNRE the likelihood function is learned from the data itself, which adds more flexibility and may allow to capture features beyond the usual assumed likelihood functions (Poisson, Gaussian, etc.). However, as the likelihood-to-evidence ratio is estimated from simulated data, which includes statistical fluctuations, the finite size of the dataset may cause small fluctuations in the final posterior calculation. This kind of fluctuations that depends on the training dataset and on the algorithm hyperparameters are common in machine learning procedures, and may be more relevant at the border of the parameter space.

Another crucial characteristic to compare is the computational cost of both approaches. For each benchmark point considered in this work the traditional MCMC analysis took $\sim 2$ days of computation in a normal laptop. On the other hand, while the TMNRE method also needs $\sim 2$ days to generate the required dataset to train the ML model, it only takes $\sim 10$ seconds to fully estimate the required posteriors once the algorithm is already trained (the training procedure required $\sim$ 30 minutes and had to be done only once). This computational speed-up opens the window to analyse an otherwise computationally prohibitive number of benchmark points. In addition, it also makes possible to test different data representations and/or experiments in a reasonable amount of time since it is trivial to combine different analyses in the TMNRE context (see \cref{sec:combAnalysis}). On the contrary, in an MCMC analysis one has to assume a new combined likelihood (that can be the multiplication of the individual likelihoods) and re-run all the chains, which would also take a few days of computation per benchmark point.

\begin{figure}[t!]
    \centering
    \includegraphics[width=1.0\linewidth]{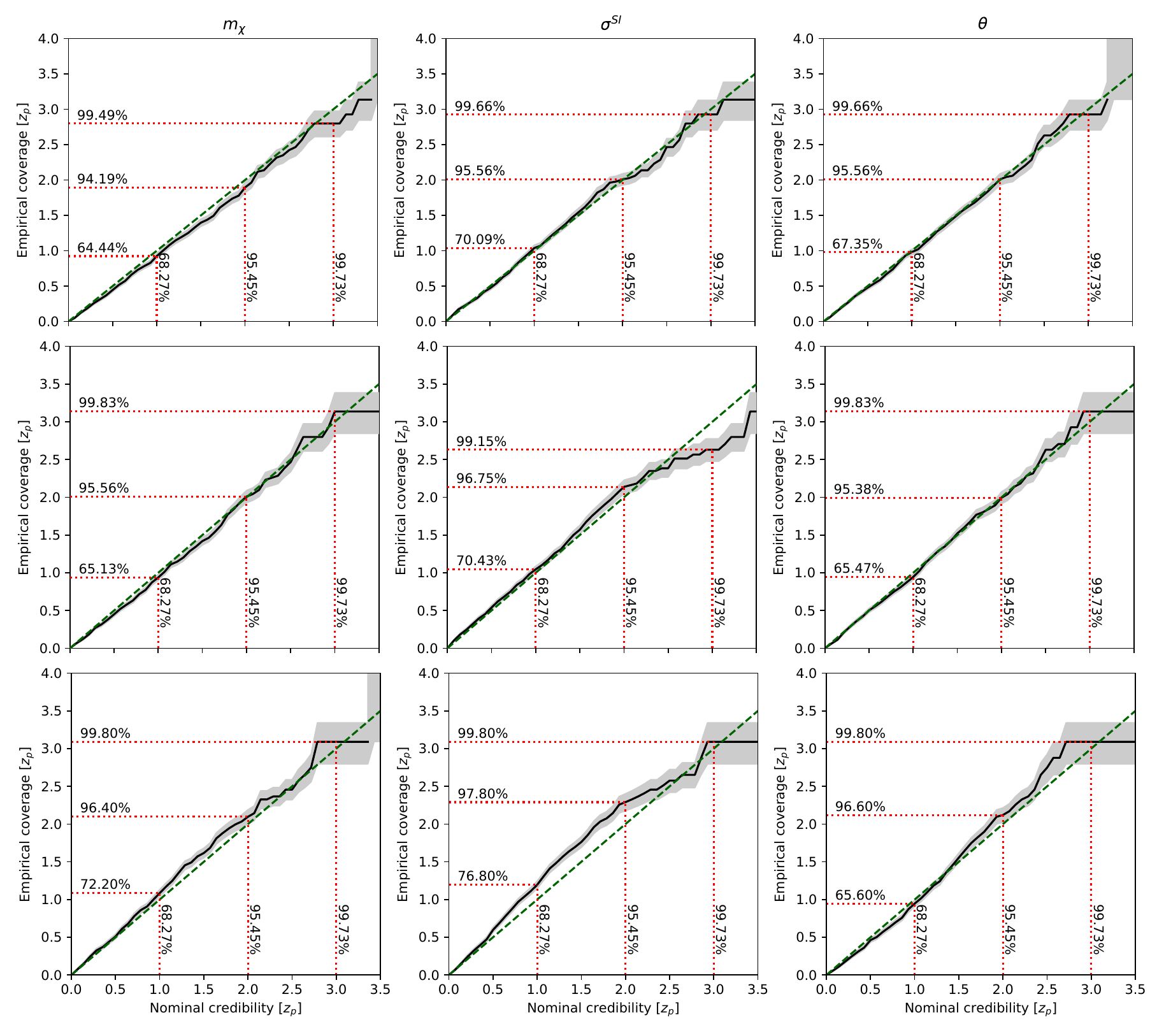}
    \caption{Empirical coverage of the posterior distributions as a function of the nominal credibility region for different values of $z_{p}=1 - \alpha$. Left, middle and right columns correspond to $m_{\chi}$, $\sigma^{\textit{SI}}$ and $\theta$ respectively. Upper, middle and lower rows correspond to total rate, differential rate and cS1-cS2 respectively.}
    \label{fig:coverage}
\end{figure}

As an additional consistency check, we compute the empirical coverage of the posterior distributions.
Although there are alternative ways of computing the coverages \cite{calibration}, we follow the procedure described in \cite{fast} that is already implemented in the \texttt{SWYFT} package.
The idea behind this procedure is that, when randomly drawing observations from $\obs, \param \sim p(\obs|\param)p(\param)$, it is expected that the true
parameters $\param$ will fall outside the $1-\alpha$ confidence region only in the $\alpha$\% of the cases. Any departure from this implies that the posterior distributions are over-confident (its empirical coverage is lower than the nominal credibility, for example if a 95\% region is constructed, but in simulations it only covers the true parameter 90\% of the time) or conservative (its empirical coverage is higher than the nominal credibility, for example if the 95\% region actually covers the true parameter 98\% of the time).
As estimating the posterior distributions may be very expensive with traditional methods, these kinds of tests are usually impossible to do.
In contrast, once trained, the TMNRE method allows for very fast estimation of posterior distributions, enabling further coverage checks. 
In \cref{fig:coverage}, we show the empirical coverage as a function of the nominal credibility region for different values of $z_{p}=1 - \alpha$. It can be seen that the estimated posteriors of all the parameters (left, middle, and right columns correspond to $m_{\chi}$, $\sigma^{\textit{SI}}$ and $\theta$, respectively) and all the data representations (upper, middle, and lower rows correspond to total rate, differential rate and cS1-cS2, respectively) are well calibrated. The only noticeable exceptions are an over-confident region (for $z_{p} \gtrapprox 3$) in the $\sigma^{\textit{SI}}$ posterior when using the differential rate, and a conservative region ($0.5 \gtrapprox z_{p} \gtrapprox 2.5$) in the $\sigma^{\textit{SI}}$ posterior when using cS1-cS2 data.

\begin{figure}[t!]
    \centering
    \includegraphics[width=\textwidth]{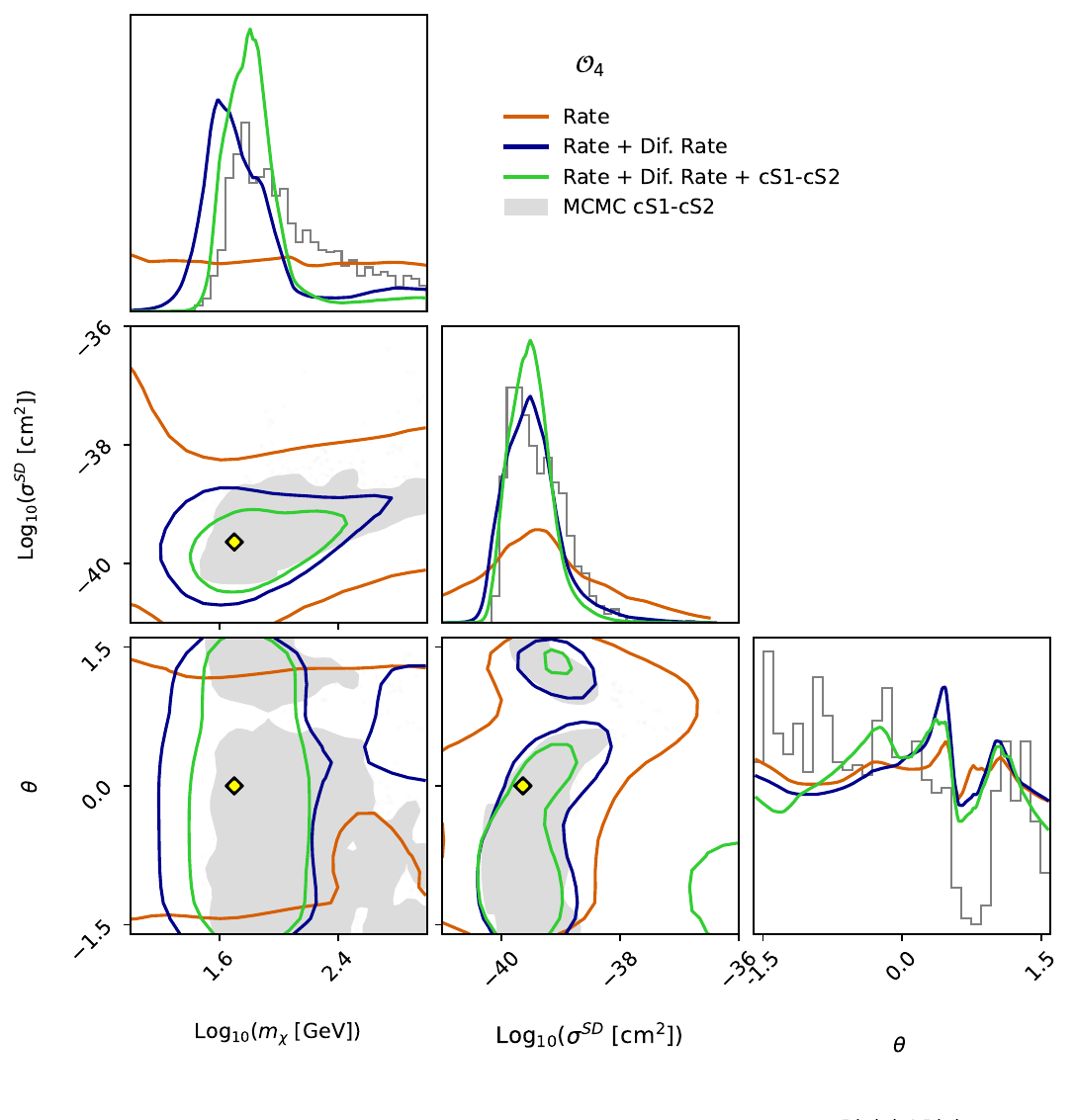}
    \caption{$90\%$ probability regions of the posterior distributions for a DM particle with an $\mathcal{O}_{4}$ interaction with $m_{\chi} = 50$ GeV, $\sigma^{SD} = 2.3\times10^{-40}$ cm$^2$, and $\theta = 0$. Colours and labels are the same as in \cref{fig:2d_posterior}.}
    \label{fig:2d_posterior_o4}
\end{figure}

In \cref{fig:2d_posterior_o4} we show the $90\%$ probability regions obtained for a DM particle with $\mathcal{O}_{4}$ interaction, $m_{\chi} = 50$ GeV, $\sigma^{SD} = 2.3\times10^{-40}$ cm$^2$, and $\theta = 0$. As in the $\mathcal{O}_{1}$ case, the TMNRE and MCMC methods provide consistent results, and both recover the true values of the analysed parameters.

Finally, notice that the total rate and the differential rate information are already contained into the cS1--cS2 representation (both are obtained as resumed features). Therefore one would expect the same results using the combination of all three data types versus using only cS1--cS2 (see Appendix~\ref{sec:mcmc} for a comparison between these two approaches). However, due to numerical and computational limitations (finite dataset, size of the networks, etc) it is a well-known fact that machine learning algorithms can benefit if some summary features are provided as inputs. In our case, it is convenient to overcome this issue by combining the individual analyses of different representations, since the procedure is straight forward and fast.

As anticipated, employing cS1--cS2 data results in a more accurate parameter estimation. For this reason, in the next section we are presenting the results only using the combination of all representations.

\subsection{Reconstruction of benchmark points} \label{sec:detectability}

It is useful to formulate a criterion that determines whether the parameters of a given benchmark point can be fully reconstructed. To do this, we consider each parameter individually and work with the 1D marginal posterior distribution, $P(\Theta_i|\obs)$, that results from the TMNRE analysis. Ideally, we want to construct a quantity that is maximal when the posterior distribution is simultaneously bounded from above and below. Generically, we can define
\begin{equation}\label{eq:int_prob_sigma}
    \mathcal{P}_{\Theta_i} = \int_{\Theta_{i,\text{min}}}^{\Theta_{i,\text{max}}} P(\Theta_i|\obs) \, d\Theta_i \; ,
\end{equation}
where the upper and lower limits of the integral are chosen, depending on the corresponding parameter, to represent either compatibility with the null signal or a parameter value from which the physics does not change anymore. We demand $\mathcal{P}_{\Theta_i} > 0.9$ as the condition for parameter $\Theta_i$ to be reconstructed, which represents a 90\% credible interval\footnote{Notice that with this definition we are only interested in imposing an upper and lower limit but the reconstruction needs not be good, i.e., we can have a large uncertainty in a reconstructed parameter.}.

For example, if we concentrate on the cross section for a given EFT operator, it seems reasonable to define $\mathcal{P}_{\sigma}$ by choosing as the lower limit in \cref{eq:int_prob_sigma} a value small enough to give $\sim 0$ events (thus compatible with no DM signal). Since the signal increases monotonically with $\sigma$, the upper limit of the integral is not relevant: for a sufficiently high value of the cross section, the posterior distribution will approach zero and the parameter is naturally bounded from above. In our analysis, we have chosen $\sigma_{\text{min}}^{SI}=10^{-49}$~cm$^{2}$ ($\sigma_{\text{min}}^{SD}=10^{-42}$~cm$^{2}$) for operators $\mathcal{O}_{1}$ ($\mathcal{O}_{4}$). For the DM mass, we need to define lower and upper integration limits, since the posterior can be non negligible for both low and high masses. These values are specific to the parameter range and to the sensitivity of the experiment we are studying. To define $\mathcal{P}_{m_{\chi}}$ we have chosen $m_{\chi,\text{min}}= 10$~GeV and $m_{\chi,\text{max}}= 400$~GeV (in general, the spectrum for larger masses is very flat and insensitive to the DM mass unless one increases the region of interest in the energy range of the region of interest \cite{Bozorgnia:2018jep}). Alternatively, if we set $m_{\chi,\text{max}} = \infty$, we could use equation \cref{eq:int_prob_sigma} to determine whether or not a lower bound on the DM mass can be extracted. This definition is denoted $\mathcal{P}_{m_{\chi}}^{\text{low}}$.

\begin{figure}[t!]
    \centering
    \includegraphics[width=1.0\textwidth]{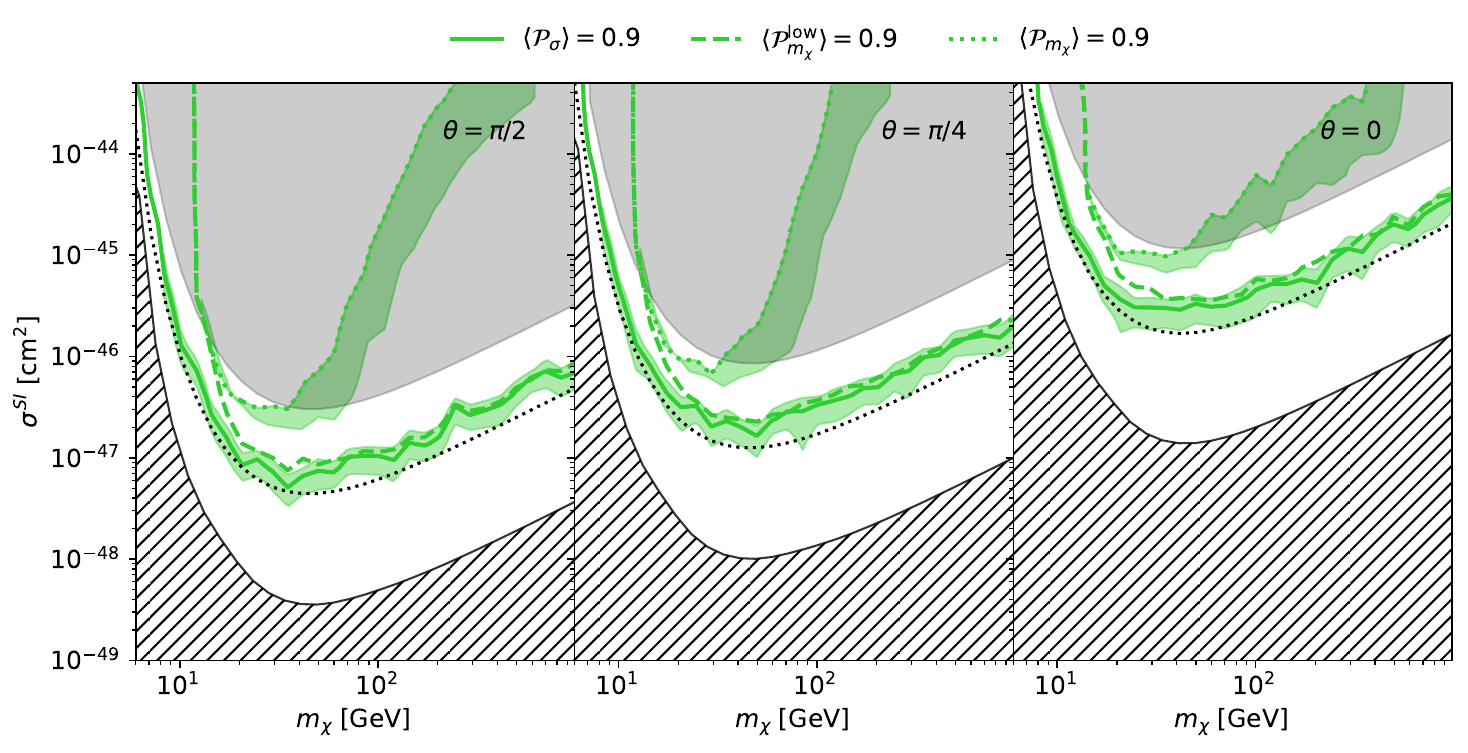}
    \caption{$\langle \mathcal{P}_{\sigma} \rangle = 0.9$, $\langle \mathcal{P}_{m_{\chi}} \rangle = 0.9$ and $\langle \mathcal{P}_{m_{\chi}}^{\rm{low}} \rangle = 0.9$ levels depicting regions where XENONnT 20~ton~yr exposure will be able to determine the DM cross section and DM mass for the $\mathcal{O}_{1}$ operator. For the $\langle \mathcal{P}_{m_{\chi}} \rangle$ case, the dashed line denotes the regions where a $m_{\chi}$ lower limit can be determined, while in the area above the dotted curves a fully measurement of $m_{\chi}$ can be achieved. Left, middle and right panels show the $(m_\chi, \sigma^{SI})$ space when fixing $\theta=\pi/2$, $\theta = \pi/4$ and $\theta=0$. The shaded grey area represents the expected exclusion region for 1.1~ton~yr, matching the current exposure presented by XENONnT~\cite{XENON:2023cxc}, and the dotted black curves the projected sensitivity for 20~ton~yr exposure, both computed with our simulator. Finally, the hatched areas show the 1-neutrino floor, following the procedure of Ref.~\cite{Billard:2013qya}.}
    \label{fig:int_prob_sigma}
\end{figure}

\begin{figure}[t!]
    \centering
    \includegraphics[width=\textwidth]{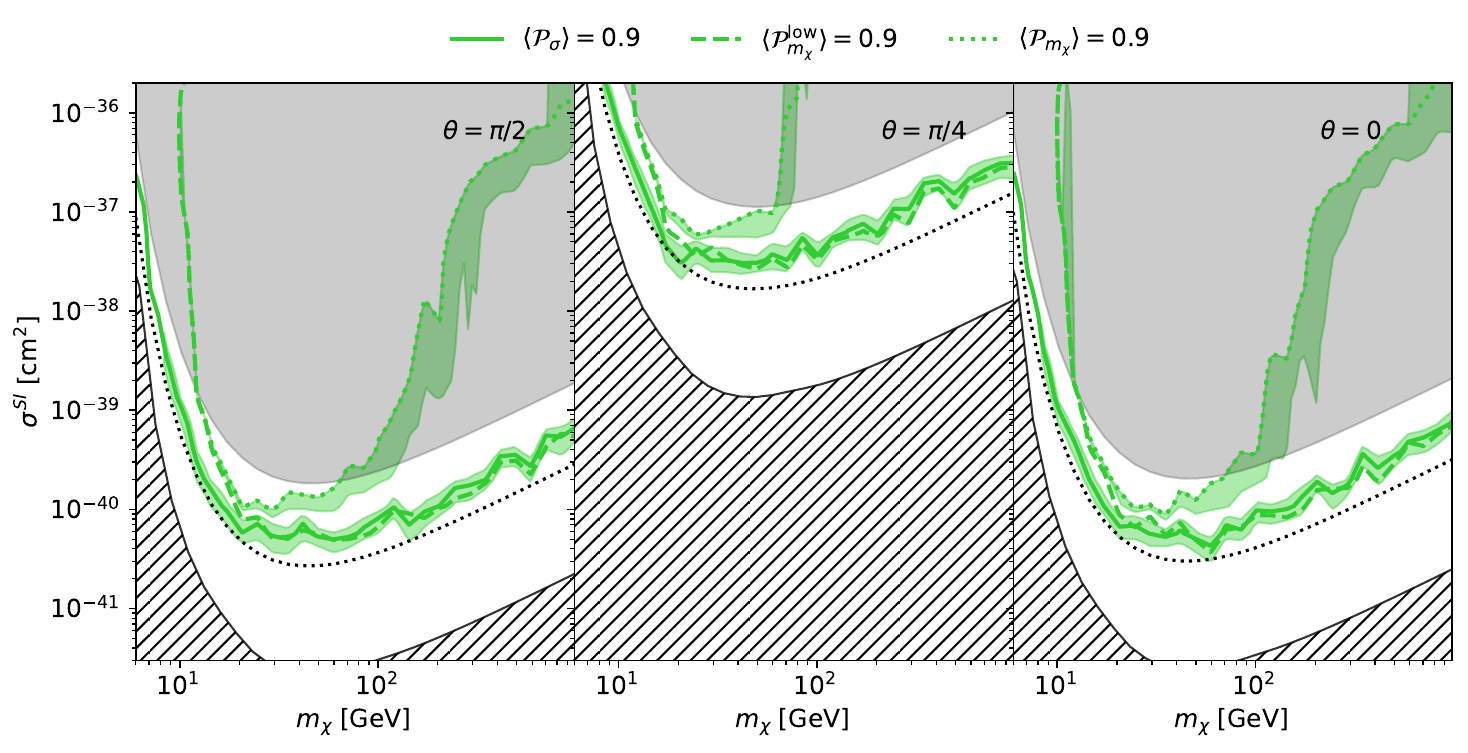}
    \caption{Same as \cref{fig:int_prob_sigma}, but for the $\mathcal{O}_{4}$ operator space.}
    \label{fig:O4}
\end{figure}

Using these criteria, we have studied the regions where the DM parameters can be reconstructed in a future xenon experiment. The results are shown in \cref{fig:int_prob_sigma,fig:O4} where, using new independent datasets, we have analysed ($m_{\chi}$, $\sigma$) slices with fixed values of $\theta$, for $\mathcal{O}_{1}$ and $\mathcal{O}_{4}$ operators, respectively. We have divided each slice in a $30 \times 30$ grid and for each point we generated five synthetic observations with the same parameters (since our simulator includes statistical fluctuations, those five synthetic observations turn out to be different from each other). We have analysed each synthetic observation with the already trained neural network (see \cref{sec:training}) and obtained the $1$D marginal posterior $P^j(\Theta_i|\obs^j)$, which we use to calculate $\mathcal{P}_{\Theta_i}^j$, where $j=1$ to $5$ represents each synthetic observation. Finally, we have applied the mentioned criteria to the mean of synthetic observation results, i.e., $\langle \mathcal{P}_{\Theta_i} \rangle = \sum_{j=1}^{5} \mathcal{P}_{\Theta_i}^j/5 > 0.9$.
It is important to clarify that each of the five synthetic observations corresponds to the expected real data if DM has the specific parameters under analysis. 
In this way, the reconstruction regions take into account possible statistical fluctuations of the data.

We should stress that, in order to carry out this analysis, we have computed the full posterior distribution and $\mathcal{P}_{\Theta_i}$ of each synthetic observations. This amounts to $13500$ realisations per operator to cover the entire parameter space. This would be impossible to do with traditional MCMC methods, as each posterior computation with its corresponding reconstruction analysis would take $\sim O($days) on a normal laptop, while with the TMNRE method takes only a few seconds once the algorithm is trained.

In \cref{fig:int_prob_sigma,fig:O4} the regions above the solid green lines correspond to points in the parameter space where the scattering cross section ($\sigma^{SI}$ in \cref{fig:int_prob_sigma} or $\sigma^{SD}$ in \cref{fig:O4}) can be reconstructed in a future xenon detector. The sensitivity can vary up to two orders of magnitude for different values of $\theta$, due to the presence of accidental cancellations between the proton and neutron contributions (also referred to as blind spots in the literature) in the calculation of the total rate. For the spin-independent operator $\mathcal{O}_{1}$ the cancellation occurs near $\theta\sim0$, whereas for $\mathcal{O}_{4}$, it is located at $\theta\sim\pi/4$.

In green dashed lines, we depict the regions where we can set a lower bound on $m_{\chi}$, which approximately coincides with the region where we can reconstruct the cross-section value (except for $m_{\chi}\lesssim 20$ GeV). Since this region is also similar to the expected exclusion limit curves, we can conclude that the projected exclusion curves for a xenon direct detection experiment provide a good reference for the region where both the cross section would be reconstructed and where a lower limit for the DM mass can be established.

The region where we are able to fully reconstruct the DM mass is delimited by green dotted curves, completely inside the region where the cross section can be reconstructed. It can be seen that we obtain a much smaller region of the parameter space in which both $m_{\chi}$ and $\sigma$ can be recovered. To establish an upper limit on $m_{\chi}$ is a more challenging problem, as for $m_{\chi} \gtrsim 50$ GeV, increasing both the DM mass and cross section particles can result in an approximately constant number of expected signal events. Finally, notice that the region in which we are able to fully reconstruct the mass and cross section is almost completely excluded by current $1.1$~ton~yr exposure results (recall that we are analysing with our Bayesian analysis the projections for 20~ton~yr exposure). If we want to reconstruct both parameters, at least in a small but non-excluded region, extra information is needed, making the complementarity with other experiments crucial and for that the procedure described in this work is ideal.

The selection of the integration limits of \cref{eq:int_prob_sigma} is not unique. This is specially relevant in the DM mass case, for which there is no zero-signal limit (contrarywise, for the cross-section analysis, the zero-signal case can be obtained for small enough $\sigma$). However, this analysis provides an initial insight into the regions of the parameter space where these properties could be measured. In order to test the impact of varying the integration limits, we performed the same analysis varying the limits in the ranges  $\sigma^{SI}_{min}\in[10^{-49.3},10^{-48.7}]$~cm$^{2}$, $\sigma^{SD}_{min} \in [10^{-42.3}, 10^{-41.7}]$~cm$^{2}$, and $m_{\chi,max}\in[400, 650]$~GeV. When further increasing $m_{\chi,max}$, we approach to the borders of our training set and therefore it is equivalent to the $m_{\chi,max} = \infty$ case (dashed green lines). The results obtained when varying the integration limits are displayed as green shaded bands around the corresponding lines on \cref{fig:int_prob_sigma,fig:O4}.

As reference, we also include the expected exclusion limit curves for 1.1~ton~yr (shaded grey area) and 20~ton~yr exposure (dotted black curve). To define these exclusion limits, we have performed a raster scan~\cite{raster} over $m_{\chi}$: for a fixed DM mass we have computed the exclusion significance, $Z$, using a binned-likelihood approach, for which we need to calculate with our simulator the expected number of signal and background (including all sources) events for a given exposure. Afterwards, we have determined the expected exclusion limits at 90\% C.L. as the cross-section value that gives $Z=1.64$. Notice that the usual spin-independent WIMP-nucleon cross section considered by the XENON collaboration corresponds to the operator $\mathcal{O}_1$ with $\theta=\pi/2$, therefore we can only compare the exclusion limits shown on the left panel of \cref{fig:int_prob_sigma} with the published ones. Then, the shaded grey area on the left panel is compatible with the current 1.1~ton~yr exclusion limit reported in Ref.~\cite{XENON:2023cxc}. On the other hand, the dotted black curve that corresponds to the projected 20~ton~yr exposure limit is less stringent than the one presented in the older Ref.~\cite{XENON:2020kmp}, since we have considered in our simulations the current background values and characteristics achieved by the collaboration in their latest publications. We expect that the sensitivity would increase to match the projections with lower background levels.

We would like to highlight that although some of our reconstruction curves have a similar shape to the exclusion limits, both approaches have different and complementary interpretations. While the latter show regions where the DM hypothesis is excluded if the experimental measurement is compatible with the background only scenario within the frequentist framework, our results indicate the regions where the mass and the couplings can be reconstructed through a full Bayesian analysis of the expected data, which also implies a positive DM detection.

This example illustrates the great advantage of the TMNRE method when combining different datasets as shown in \cref{sec:combAnalysis}. In practise, this would also be applicable to the joint analysis of data from distinct direct detection experiments. In such a case, the analysis for each detector would be performed separately and the combined result would be obtained following the strategy described in this article.


\section{Conclusions}\label{sec:conclusions}

In this article, we have applied state-of-the-art machine learning methodologies to perform a Bayesian analysis of dark matter direct detection data. In particular, we have assessed the performance of the Truncated Marginal Neural Ratio Estimation method, a Simulation Based Inference technique, to estimate particle model parameters. Unlike in traditional Monte Carlo Markov Chain methods, there is no need to specify the likelihood functional form, which instead it is estimated from the simulated data. We have demonstrated that datasets of different kinds, which have been trained individually to analyse the same parameter space, can be combined in a simple and fast way. This allows the usage of results from different experimental facilities in a modular fashion, without the need to re-train the machine-learning algorithms or define a combined likelihood. Additionally, marginal distributions can be computed automatically, considerably speeding up the computation if we are only interested in a subset of parameters.

As an application,  we have analysed the potential of a future xenon detector, based on XENONnT, to reconstruct the DM parameters (mass and couplings). For some benchmark points we have validated our results with traditional MCMC methods. Although both procedures provide similar and compatible posteriors, the algorithm presented in this work turns out to be computationally very efficient and orders of magnitude faster than MCMC methods. This makes it possible to analyse large parameter spaces and to use it in applications which require a large number of evaluations. As an example of the latter, we have employed TMNRE to determine the regions of the parameter space where the DM mass and spin-dependent and independent couplings to protons and neutrons can be reconstructed.

To illustrate how advantageous the statistical analysis presented in \cref{sec:combAnalysis} is when combining datasets, we have used three different representations: total event counts, differential event rates, and cS1--cS2 distributions. As expected, our results indicate that the combination of all representations leads to better parameter recovery and coverage of the DM parameter space, specially thanks to the inclusion of cS1--cS2. In \cref{fig:int_prob_sigma} and \cref{fig:O4} for $\mathcal{O}_{1}$ and $\mathcal{O}_{4}$ operators, respectively, we have presented the regions where $m_{\chi}$ and/or $\sigma$ can be estimated (the value of $\theta$ can not be reconstructed) for the full exposure of XENONnT if a positive DM signal is detected. An observation in a single experiment would be insufficient to fully remove degeneracies in the DM parameter space and complementary observations with other targets would be necessary. The technique presented in this article is ideal to combine future experimental results with independently-trained Machine-Learning models.

Finally, we have developed a Python library, \texttt{CADDENA}, which performs a modular Bayesian analysis of direct detection experiments. This package includes all the material used in this work, pre-trained models and simulation data, which allows the user to extend the analysis to new DM interaction models and setups.

\vspace{2.5mm}
\paragraph{Acknowledgments.}

ADP, DC and MdlR acknowledge support from the Comunidad Autonoma de Madrid and Universidad Autonoma de Madrid under grant SI2/PBG/2020-00005, and by the Spanish Agencia Estatal de Investigaci\'on through the grants PID2021-125331NB-I00 and CEX2020-001007-S, funded by MCIN/AEI/10.13039/501100011033. DC also acknowledges support from the Spanish Ministerio de Ciencia e Innovaci\'on under grant CNS2022-135702.

\section*{Appendix}
\appendix

\section{MCMC sampling method} \label{sec:mcmc}

To validate the TMNRE method, we have performed, for two benchmark points, a Bayesian analysis employing \texttt{MultiNest}~\cite{Feroz:2007kg,Feroz:2008xx,Feroz:2013hea}, an MCMC nested sampling procedure,  using the Python implementation \texttt{bilby}~\cite{bilby_paper}.

We considered the profiled log-likelihood-ratio test statistic,
\begin{equation}
    q(\param) = -2 \, \ln \left( \frac{\mathcal{L}(\param)}{\mathcal{L}(\hat{\param})} \right) ,
\end{equation}
where $\mathcal{L}$ is the likelihood function that describes the data given the model parameters, and $\hat{\param}$ indicate the quantities that maximise the likelihood.

For the analysis taking into account the total and differential rate, we have employed a binned treatment and we have assumed that the number of counts in each bin follows a Poisson distribution. Thus,
\begin{equation}
    q(\param) = 2 \left[ \sum_{k=1}^{n_{bins}} \, N^k_{th}(\param) \, - \, N^k_{\rm obs} \, + \, N^k_{\rm obs} \, \ln \left( \frac{N^k_{\rm obs}}{N^k_{th}(\param)} \right) \right],
\end{equation}
with $N^k_{\rm obs}$ and $N^k_{th}(\param)$ the number of observed and expected events in the bin $k$. We highlight that $N^k_{th}(\param)$ takes into account both signal and background contributions. $n_{bins}$ is the number of total bins, equal to 1 for the total rate analysis while for the differential rate we have considered a bin width of $1$ keV$_{\rm NR}$ and a range (3, 61) keV$_{\rm NR}$.

For the cS1--cS2 space we considered 97 bins in cS1 within (3, 100) PE, and 70 bins in cS2 within (100, 10000) PE. However, since many of those bins can be empty, we model the log-likelihood-ratio test statistic with a mean square error distribution
\begin{equation}
    q(\param) = \frac{1}{n_{bins}} \sum_{k=1}^{n_{bins}} \, (N^k_{th}(\param) \, - \, N^k_{\rm obs})^2.
\end{equation}

\subsection{MCMC vs TMNRE}

\begin{figure}[t!]
    \centering
    \includegraphics[width=0.55\linewidth]{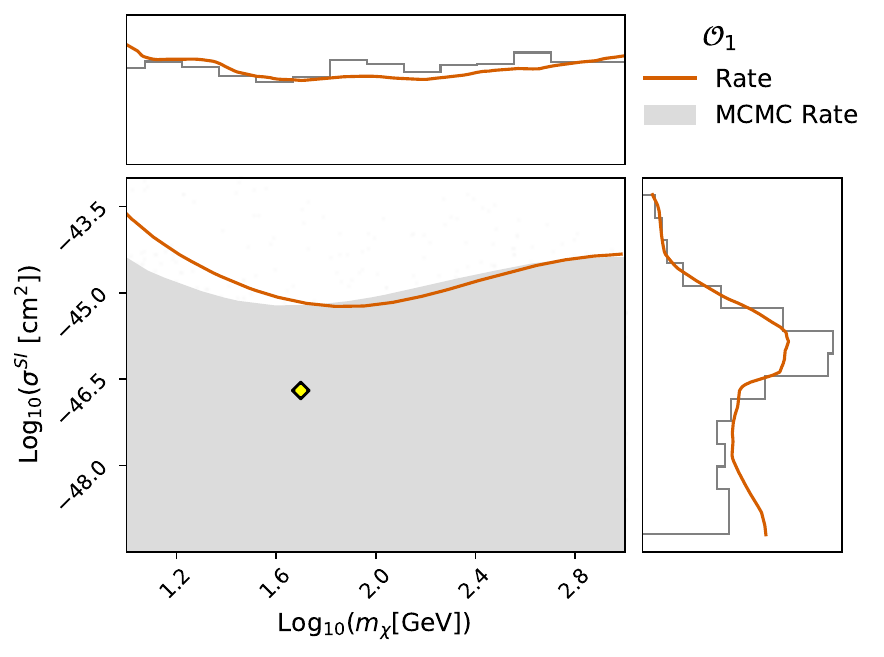}
    \includegraphics[width=0.55\linewidth]{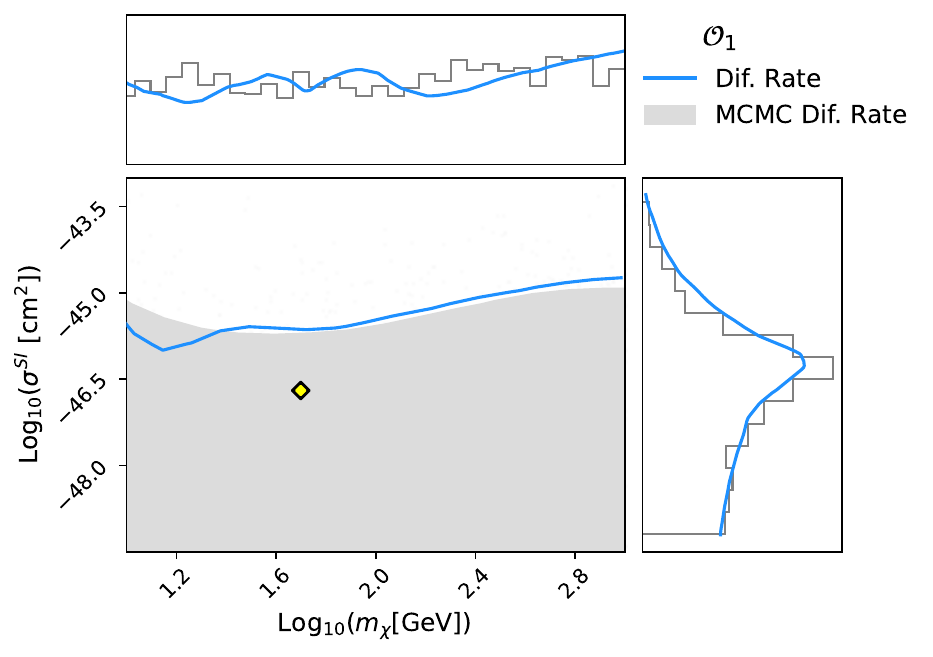}
    \includegraphics[width=0.55\linewidth]{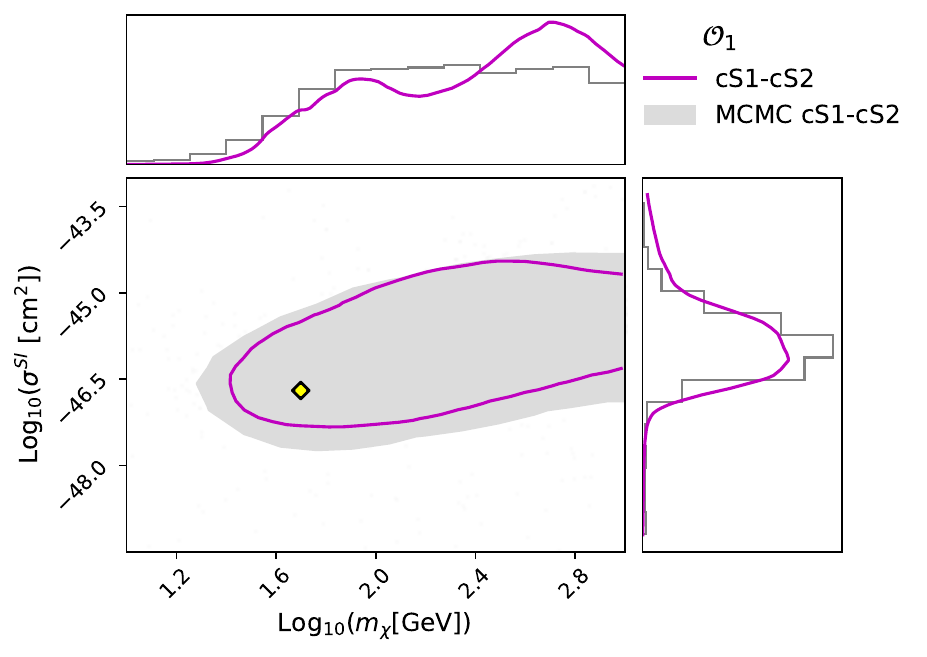}
    \caption{$2$D $(m_\chi, \sigma^{SI})$ marginal posterior distributions and their corresponding $1$D marginal posteriors for a dark matter particle with an $\mathcal{O}_{1}$ interaction, $m_{\chi} = 50$ GeV, $\sigma^{SI} = 2\times10^{-47}$ cm$^2$, and $\theta = \pi/2$. In solid coloured curves we show the posteriors obtained with our procedure while in grey with MCMC.
    }
    \label{fig:MCMCvsSWYFT}
\end{figure}
In \cref{fig:MCMCvsSWYFT} we show a comparison between the $90\%$ probability regions of the posterior distributions obtained with our method using TMNRE and with MCMC. The example considers $\mathcal{O}_{1}$ interaction, $m_{\chi} = 50$ GeV, $\sigma^{SI} = 2\times10^{-47}$ cm$^2$, and $\theta = \pi/2$, and we compare the results for the three data representations mentioned above. We can see that all $1$D and $2$D marginal posterior obtained with both methods are compatible.

\begin{figure}[t!]
    \centering
    \includegraphics[width=\textwidth]{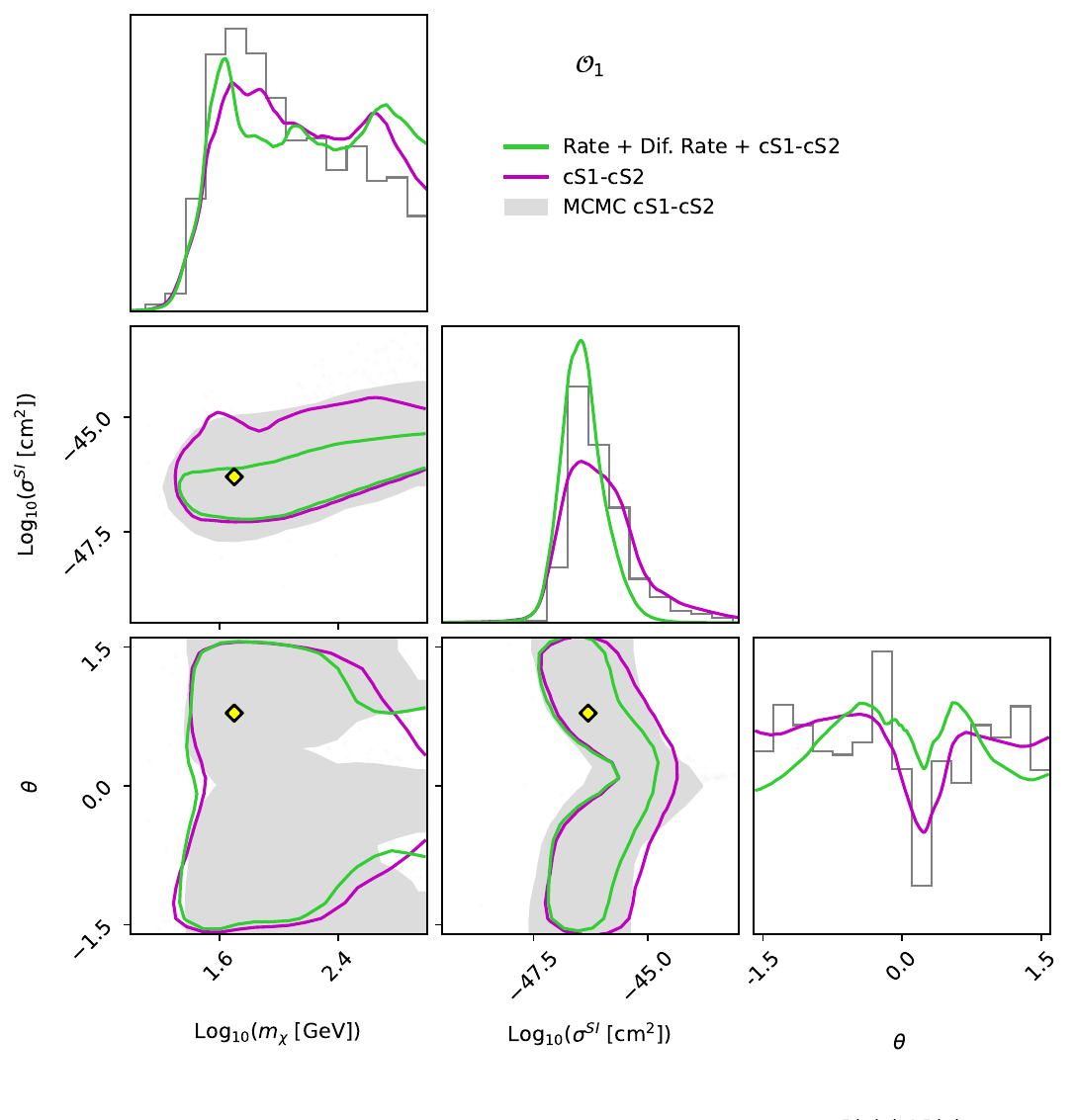}
    \caption{Posterior distributions for a dark matter particle with an $\mathcal{O}_{1}$ interaction, $m_{\chi} = 50$ GeV, $\sigma^ {SI} = 2\times10^{-47}$ cm$^2$, and $\theta = \pi/4$. In green we show the results obtained analysing total number rate + differential rate + full cS1--cS2 information, while in magenta only the full cS1--cS2 representation is used. In grey the results with a traditional MCMC method are presented.}
    \label{fig:2d_posterior_nocomb}
\end{figure}

Finally, in \cref{fig:2d_posterior_nocomb} we can see the impact of combining different data representations, for $m_{\chi} = 50$ GeV, $\sigma^{SI} = 2\times10^{-47}$ cm$^2$, and $\theta = \pi/4$. In particular, in the $1$D and $2$D marginal posteriors involving $m_{\chi}$ and $\sigma^{SI}$ the benefits of combining data is noticeable. On the other hand in the $1$D $\theta$ marginal posterior an artificial tendency pointing toward a wrong parameter value is present for the combined case (although the posterior is significant in the entire parameter space), which is absent using only cS1--cS2 information or in the MCMC realisation. This behaviour in inherited from the analysis using only the total event rate, representation that can not provide enough information to the algorithm to learn the subtleties of the $\theta$ parameter that is highly degenerate. In that regard we would like to mention that one could think that there should be no difference when using only cS1--cS2 or the combination of all representations since both the rate and the differential rate are summary features of cS1--cS2. Even though in principle this is true for an infinitely large network and dataset, due to numerical and computational limitations it is well known that machine learning algorithms can benefit if some summary features are provided as inputs.

\section{\texttt{CADDENA}: A Python implementation} \label{sec:caddena}

We have developed a Python implementation for the modular Bayesian analysis of dark matter direct detection experiments.
This library is publicly available on github~\footnote{\href{https://github.com/Martindelosrios/CADDENA}{https://github.com/Martindelosrios/CADDENA}} and can be easily installed in any Python environment.

The main function of the package can be used to estimate the likelihood-to-evidence ratio for a given realisation,
\begin{lstlisting}[language=Python]
from CADDENA import caddena
caddena.ratio_estimation([realisations_list], pars_ar, [models_list])
\end{lstlisting}
where \texttt{realisations\_list} is a Python list with the data that will be analysed, \texttt{pars\_ar} is a numpy array with the parameters that will be paired with the data (see \cref{sec:bayanalysis}) and \texttt{model\_list} is the list of models that will be used to analyse the data. Here is important to remark that each model in \texttt{model\_list} should be consistently trained to analyse each item in \texttt{realisations\_list} and that all the models should be trained to predict the same parameters in order to be combined as explained in \cref{sec:combAnalysis}.

In the current version, all the data and the neural networks models used in this work are available without the need of re-training. This includes the simulations and the neural networks weights of the models for DM with $\mathcal{O}_{1}$
and $\mathcal{O}_{4}$ interactions for the three data representations considered.
All these features allows the reproduction of the results presented in this paper.
In the near future we plan to incorporate more models to analyse the complete set of NR-EFT operators and the response of other experimental facilities.
All the available models can be listed and loaded with the following commands:

\begin{lstlisting}[language=Python]
# Listing available models
from CADDENA import caddena
caddena.Model.available_models()

# Loading models
from CADDENA import models
models.ModelName

# Loading the pre-trained weights
models.ModelName.load_weights()

# Check is a model is already trained and ready to be used
models.ModelName.trained_flag()
\end{lstlisting}

The package also allows to extend the current analysis in a simple way. The inclusion of new experiments or/and data representations can be done through the creation of custom models with
\begin{lstlisting}[language=Python]
from CADDENA import caddena
ModelName = caddena.Model(
    network,
    trainer,
    path_to_weights,
    test_data,
    comments
)
\end{lstlisting}
where \texttt{network} and \texttt{trainer} are \texttt{SWYFT} objects containing the corresponding machine learning models\footnote{For more information on how to create the corresponding \texttt{SWYFT} objects, please refer to \href{https://swyft.readthedocs.io/en/stable/}{https://swyft.readthedocs.io/en/stable/}.}, \texttt{path\_to\_weights} and \texttt{test\_data} are optional inputs in case the models were already trained and \texttt{comments} is also an optional input to add some extra information that will be printed when calling the model.

\bibliographystyle{JHEP} 
\bibliography{main-v2.bbl}

\end{document}